\documentclass[preprint,aps,draft]{revtex4}

\usepackage{graphicx}
\usepackage{dcolumn}
\usepackage{bm}
\setkeys{Gin}{draft=false}

\begin{document}


\title{Wavelet analysis of magnetic turbulence in the Earth's plasma sheet}

\author{Z. V\"{o}r\"{o}s}
\email{zoltan.voeroes@oeaw.ac.at}
\author{W. Baumjohann}%
\author{R. Nakamura}
\author{A. Runov}
\author{M. Volwerk}
\altaffiliation[Also at ]{Max-Planck Institute for Extraterrestrial Physics, Garching, Germany}
\author{T.L. Zhang} 
\affiliation{%
Space Research Institute, Austrian Academy of Sciences, Graz, Austria
}%

\author{A. Balogh}
\affiliation{
Imperial College, London, UK
}%

\date{\today}

\begin{abstract}
Recent studies provide evidence for the multi-scale nature of magnetic turbulence in the plasma sheet. 
Wavelet methods represent modern time series analysis techniques suitable for the description of 
statistical characteristics of multi-scale turbulence. Cluster FGM (fluxgate magnetometer) magnetic field high-resolution 
(~67 Hz) measurements are studied during an interval in which the spacecraft  are in the plasma sheet. 
As Cluster passes through different plasma regions, physical processes exhibit non-steady properties 
on magnetohydrodynamic (MHD) and small, possibly kinetic scales. As a consequence, the implementation of  
wavelet-based techniques becomes complicated due to the statistically transitory properties of 
magnetic fluctuations and finite size effects. Using a supervised multi-scale technique which allows 
existence test of moments, the robustness of higher-order statistics is investigated. 
On this basis the properties of magnetic turbulence are investigated for changing thickness of the 
plasma sheet.
\end{abstract}

\maketitle

\section{\label{sec:level1}INTRODUCTION}

Direct observations of the velocity and magnetic field in the plasma sheet have revealed strong 
intermittent fluctuations in the temporal and spatial domains. These observations were attributed to
turbulence \cite{bor97}.
Eddy turbulence rather than Alfv\'{e}nic turbulence seems to prevail  and the most important dissipation 
mechanisms include a multi-scale cascade of  energy to non-magnetohydrodynamic (non-MHD) 
scales and an electrical coupling of 
the turbulent flows to the ionosphere \cite{bor03}.  
In contrast with the classical hydrodynamic or MHD homogeneous  
turbulence picture \cite{frisch, biskamp}, 
MHD turbulence in the plasma sheet  is not free from boundary effects \cite{bor03, volw}. 
Moreover, due to the movement of  boundaries (e.g. the plasma sheet boundary layer (PSBL), or a flow channel 
built up during rapid plasma flows) and  the transitory character of driving mechanisms 
(e.g. shear flows, sporadic reconnection, MHD instabilities), the observed processes show intermittence 
in time rather than intermittence in space, an important clue for distinguishing  homogeneous 
and non-homogeneous plasma sheet flows \cite{vor03}. Intermittence is related to long-tailed probability distributions,
hence to higher order statistical moments. In general, statistical moments are defined by the average
of the powers of a random variable.
In solar wind intermittence studies it is customary to use
$q$-th order absolute powers of velocity, magnetic field, etc. increments  (so-called $q$-th order 
structure functions) \cite{pagel, mars}, which allow to investigate the multi-scale scaling features
of fluctuations with long-tailed probability distributions. Direct studies of empirical
probability densities of increment fields in the solar wind revealed departures from a Gaussian
distribution over multiple scales \cite{sorr} and an increase of intermittence
towards small scales \cite{mars94}.  Another class of intermittence studies
uses large deviation concepts reconstructing distribution functions of burstiness of local fluctuations
in considered fields \cite{vor02, vor03}. An alternative  for studying multi-scale space plasma 
intermittence is represented by the wavelet method \cite{kov, cons}, which also proved to be useful
in resolving multi-scale cascading features of a current disruption event in the Earth's plasma 
sheet \cite{lui}. In this paper we investigate magnetic field intermittence using $q$-th order moments 
(average of the powers) of wavelet coefficients. In the following under  statistical moments we mean 
$q$-th order moments of 
wavelet coefficients.
To be consistent, we  specify the main problems related to the estimation
of moments in the plasma sheet.  
First of all, a robust estimation of higher-order statistical characteristics of plasma sheet 
turbulence  requires the processing of long time series, while the recited non-steady features might 
change the internal structure of the observed turbulence. Contrarily, if the measurements are limited 
to too short time intervals, finite size effects lead to the divergence of higher order moments, 
and the description of fluctuations  which show large deviations from a mean value becomes impossible. 
Additional difficulties are introduced by moment  estimators which  are not sensitive to 
the non-existence or divergence of  statistical moments. 
Under the circumstances, for the proper recognition of  the nature of large deviations in turbulence, 
at least three basic conditions have to be  taken into account:  1.) the motion of  PSBL, 2.)  the time 
evolution of the driving and/or dissipation mechanisms and 
3.) the number of existing statistical moments.  
One can detect PSBL motion (e.g. plasma sheet thinning or thickening) from direct, 
preferably  multi-spacecraft observations \cite{nak, nak1}. The driving and dissipation mechanisms 
are obviously
dependent on the physical process examined. In this paper we analyse bursty bulk flow (BBF)
associated magnetic fluctuations. Sporadically occurring BBFs can stir the plasma sheet plasma
very efficiently, because they are the carriers of  decisive amounts of
mass, momentum and magnetic flux \cite{angel, schod}. 
Here the driver is the plasma flow itself, while the increased
small scale power of the magnetic fluctuations can be used for a proper detection of ongoing 
dissipation processes. Then a statistical wavelet-based test ensuring the existence of moments can enhance 
the reliability of the intermittence level estimations. 

\section{\label{sec:level1}WAVELET METHODS}

In this paper we will use wavelet methods
for  the estimation of the power of small scale magnetic fluctuations, $c_f$,
and for the computation and existence test of statistical moments, as well.

It was shown in Ref. 19 that a semi-parametric wavelet technique, based on a fast pyramidal algorithm,
allows unbiased estimations of the scaling parameters $c_f$ and $\alpha$ 
in the scaling relation for power spectral density $P(f)\sim c_f f^{-\alpha}$,
where $c_f$ is a nonzero constant. The algorithm consists of several steps. First, a discrete wavelet 
transform of the time series X(t) is performed over a dyadic grid $(scale, time)=(2^j,2^j t)$ and $j,t 
\in \bf{N}$. Then, at each octave $j=log_2 2^j$, the variance $\mu_j$ of the discrete wavelet coefficients 
$d_x(j,t)$  is computed through:
\begin{equation}
\mu_j = \frac{1}{n_j}\sum_{t=1}^{n_j}d_x^2(j,t) \sim 2^{j\alpha}\ c_f
\end{equation}
where $n_j$ is the number of coefficients at octave $j$.
Finally, from Equation (1) $\alpha$ and $c_f$ can be estimated 
by constructing 
a plot of $y_j\equiv log_2 \mu_j$ versus $j$ (logscale diagram) and by using a weighted 
linear regression 
over the region  $(j_{min}, j_{max})$ where $y_j$ is assumed to be a straight line. 

Generalizing Equation (1), for a class of multifractal processes,  the so-called
partition function   can be introduced through the $q$-th order moments  
of the wavelet coefficients \cite{abry}:
\begin{equation}
\mu_j^q = \frac{1}{n_j}\sum_{t=1}^{n_j}d_x^q(j,t) \sim 2^{j(\zeta(q)+q/2)} 
\end{equation}
The partition function measures not only the scaling of the moments, but also the higher order 
dependencies of the wavelet coefficients. When $\zeta(q)$ is plotted against $q$ 
together with $95\%$ confidence intervals of the mean $\zeta(q)$
(multiscale diagram), self-similar (fractal) and multifractal processes can be distinguished 
\cite{abry}. A nonlinear $\zeta(q)$ is a signature of multifractal scaling and small scale intermittence. 
However, the  wavelet estimator may yield finite values for $\mu_j^q$ even in 
situations when a $q$-th order moment does not exist or diverges. This can happen when the higher
order dependencies of the estimator rather than the true scaling of the moments  are observed over
a range of scales.
The bias introduced by the estimator
may prevent us from discriminating between monofractal and multifractal processes, because of
the false nonlinear dependence of $\zeta(q)$ on $q$. In order to deduce the 
proper support of the partition function, $(q_{min},q_{max})$, over which all the moments exist 
and are finite, a simple method based on characteristic functions was proposed \cite{gonc}.
Here we shortly summarize this method. 
The empirical characteristic function 
for the time series $X_n$ $(n=1,...,N)$ is computed as $F(u) = N^{-1}\sum_n e^{iuX_n}$. It represents
the Fourier transform of the probability distribution of $X$. It has been proven that $F$ has 
as many continuous derivatives at $u=0$ as the probability distribution of $X$ has finite positive integer 
moments.The generalization of the local integer degree of differentiability to real-valued degrees of
differentiability is possible using the concept of H\"{o}lder regularity.  
The H\"{o}lder 
regularity of $F$ at the origin ($u=0$) controls the number of existing real-valued 
moments while $q_{max}\le2$. Only an averaged regularity of $F$ around the origin is
ensured to exist for moments exceeding 2 \cite{kawa}. It introduces  limitations to the testing procedure
of existing moments  larger than 2. It is possible, however, even in this case to estimate
the lower and upper limit for the largest existing positive moment $q_{max}$ \cite{gonc}.
Since the wavelet transform, using a wavelet $\psi$ with vanishing moments $M_{\psi}$, is well suited 
for estimating the H\"{o}lder regularity \cite{mallat}, F is wavelet transformed, and only the
wavelet coefficients at the origin $d_F(s, u=0)$ are considered further. Here instead of $j$
the notation $s$ is used for describing the characteristic scales of $F$. 
The number of vanishing moments $M_{\psi}$ allows us to cancel or decrease the effects of linear or 
polynomial trends and ensures that the wavelet details are well defined. This is because a wavelet
with $M_{\psi}$ vanishing moments is orthogonal to the polynomials of degree $M_{\psi}-1$ and the 
wavelet transform acts as a multiscale differential operator of order $M_{\psi}$ \cite{mallat}.
The H\"{o}lder regularity of a signal can be estimated by wavelets with vanishing moments 
exceeding that H\"{o}lder regularity by at least  1.
Then the H\"{o}lder regularity of $F$ can be estimated from the decay of the wavelet 
coefficients across the scales.
It allows to estimate
$q_{max}$  from a linear regression of $log_2d_F(s, u=0) $ versus $s$. The largest existing
negative moment $q_{min}$ can be estimated by applying the same procedure as above, but for the inverse 
variable $X^{-1}$ \cite{gonc}.
Further difficulties arise with finding the proper scales $s$ over which the  H\"{o}lder 
regularity of $F$ can be evaluated. We will demonstrate that three different scaling ranges of
$F$ appear. One of them reflects the scaling properties of the chosen wavelet, 
and therefore the observed scaling over that range is not related
to the pysical process itself. This scaling range can be easily identified by changing the basic
feature of the analysing wavelet: the number of vanishing moments  $M_{\psi}$. 
To this end an appropriate wavelet has to be chosen which allows  changing $M_{\psi}$. In this paper
we use  $m$-th order derivatives of the Gaussian wavelet which have $m$ vanishing moments.
The remaining two scaling
ranges reflect a symmetry property of the estimator. An exchange of the time series $X$ by $X^{-1}$ 
results in a mirroring of the scaling regimes with respect  to the characteristic scale  
which separates the tail from the body of the underlying distribution function.
This symmetry feature of the estimator allows us to obtain both $q_{min}$ and $q_{max}$ at once, evaluating
only the scaling properties of the characteristic function for the time series $X$ \cite{gonc}.
We will consider further details of this method later.

\section{\label{sec:level1}BBF associated magnetic turbulence on July 30, 2002} 
\subsection{\label{sec:level2}Event overview}
In this paper we analyse burst mode (67 Hz) magnetic data from the Cluster fluxgate magnetometer
(FGM) \cite{balogh} during the interval 1730-1900 UT on July 30, 2002, when the Cluster (C) spacecraft
were at the GSM (Geocentric Solar Magnetospheric) position (-16, -11, 2)$R_E$. 
The GSM coordinate system will be used throughout the paper, in which the $x$ axis is defined along
the line connecting the center of the Sun to the center of the Earth. The origin is defined
at the center of the Earth and is positive towards the Sun.
Figure 1a shows the $B_X$ component from C 1,3.
From 1730 to 1740 UT both spacecraft are in the lobe ($B_X\sim 30$ nT). After 1740 UT the 
spacecraft approach the neutral sheet ($B_X \to 0$) where they remain until almost 1900 UT.
The $B_Z$ component from C1,3 is depicted in Figure 1b together with a dashed line at the top
indicating the occurrence of  intermittent groups of BBF events. During the first half of the interval
the velocity of the plasma flow increases up to 1500 km/s (not shown). BBFs drive the magnetic fluctuations
of both $B_X$ and $B_Z$ components and cause a clear dipolarization of the magnetic field (increase of 
$B_Z$) at the beginning. Figure 1c shows the time evolution of the power of the $B_Z$  fluctuations,
$c_f(B_Z)$, which is estimated through Equation 1,  in the logscale diagram, at the scale $j=4$ ($\sim 0.33$ s).
$c_f(B_Z)$ is estimated within  sliding overlapping windows of width $61$ s with a time shift $4$ s.
All the variations of $c_f(B_Z)$ are relative enhacements  to the lobe values which are normalized to
$1$. In this way $c_f(B_Z)$ represents a way of quantifying the relative power of the fluctuations 
at a given scale. In the following we restrict our analysis to the subintervals $A$ and $B$, 
depicted in Figure 1c. During  interval $A$, $c_f(B_Z)$ fluctuates intermittently on both C1 and C3.
The difference in $B_X$  measured at the locations of C1 and 3 changes substantially, 
indicating spatial gradient lengths of the order of the distance between the spacecraft.
C1 and C3 are in  opposite hemispheres in a distance $\sim 4000$ km 
before 1800 UT. The vertical position to the current sheet allows to use C1, C3 magnetic observations
for rough estimation of the influence of PSBL. At the beginning of the interval A,
$B_X$ decreases from $\sim20$ nT to $\sim0$ nT showing large fluctuations about the mean value. 
After 1752 UT the fluctuations achieve $\sim-20$ nT. Both the large fluctuations and the values 
close to $-20, +20$ nT indicate that
the magnetic fluctuations during the interval A might be influenced by the PSBL. 
From the decreasing gradients after 1800 UT, we deduce that, the plasma sheet gradually 
becomes thicker. Both spacecraft stay  closer to the neutral
sheet and the amplitude of fluctuations is also considerably smaller. 
Therefore, the influence of the PSBL on turbulence characteristics might be weaker during the interval
$B$. We will compare the higher-order statistical characteristics during the two intervals using magnetic 
data from C3. However, before
that, the proper support of the partition function (Eq. 2) has to be evaluated.

\subsection{\label{sec:level2}Scaling of the characteristic function}
Figure 2 shows the scaling properties of F computed for the $B_Z$ component on C3 during  period $A$. 
The continuous line corresponds to the estimated dependence of $d_F$ on $s$ at the origin in the log-log
plot. The dashed-dotted lines show different scaling regimes.
The interpretation follows the way proposed in Ref. 20.
The maximum variance of F is controlled by the maximum value of $B_Z$. When the analysing scales
go below $s_{min}\sim 1/max(B_Z)$, the characteristic function is oversampled in the vicinity of the
origin. Below  $s_{min}$, the regularity of the analysing wavelet is observed. 
Therefore it shows a scaling $\sim s^{M_{\psi}}$, which is different from  the scaling of $F$. 
The maximum scale, $s_{max}$, which separates the tail from
the body of the underlying distribution function, can be found experimentally. For the scales $s>>s_{max}$
the same scaling is observed as  would have been obtained, if  we had analysed a 
random variable $B_Z^{-1}$ instead of $B_Z$. Figure 2 shows that for $log_2s >-3$, $d_F$ scales as $\sim s^{\rho^-}$, 
therefore
for the negative moments $q_{min} = \rho^-$. Between the scales $(s_{min}, s_{max})$, 
the characteristic function scales as $s^{\rho^+}$.
The estimated values are $\rho^-\sim-1.0\pm0.1$ and $\rho^+\sim2.2\pm0.1$. 
For  period B, $\rho^-\sim-1.0\pm0.1$ and 
$\rho^+\sim2.4\pm0.1$ (not shown).
In both cases the scaling exponent $\rho^+$ is larger than 2.
In such a case the conditions for existing moments can be formulated in terms of an averaged 
H\"{o}lder regularity of $F$ at the origin. To be able to detect the lower and upper bounds for an 
unknown average regularity,
the number of vanishing moments ($M_{\psi}$) of the analysing wavelet
has to be successively increased. First a low regularity wavelet can be chosen, e.g. the second derivative
of the Gaussian wavelet. When the scaling exponent $\rho^+$ is equal or larger than 2, as in our case above,
we can increase $M_{\psi}$ until  $\rho^+$ will achieve $M_{\psi}$ between the scales 
$(s_{min}, s_{max})$.  
It has been shown  that when  $\rho^+\sim M_{\psi}$ is obtained,
$\rho^+<q_{max}<\rho^++1$ \cite{gonc}. On this basis $\zeta(q)$ can be computed over the support 
$(q_{min},q_{max})=(-1,3)$. The same computations for the $B_X$ component of the magnetic field
also give $q_{min}\sim-1$. Because of the small separation between $s_{min}$ and $s_{max}$, however, it 
is not so straightforward to estimate $q_{max}$ for $B_X$. Our estimation based on linear regression
for different $M_q$s is $q_{max}=1.5\pm0.5$. 
In what follows, the support $(q_{min},q_{max})=(-1,3)$ will be used for both $B_X$ and $B_Z$, 
having in mind that the estimate of $\zeta(q, B_X)$ is less reliable for $q>2$.

\subsection{\label{sec:level2}Scaling of the moments}
The dependence of $\zeta(q)$ on $q\in(q_{min},q_{max})$ for $B_X$ and $B_Z$  estimated over two
different range of scales during  interval A is depicted in Figure 3. The large time scales $0.67-5.4$ s,
correspond to spatial scales of $670-5400$ km, assuming $1000$ km/s plasma flow velocities. The 
smaller value is of the order of  the proton gyroradius in the plasma sheet, while the larger value is
limited by the length of the intervals chosen. These values represent the lower end  
of the scale range of the MHD 
regime in turbulence.  Similarly, the small time scales $0.08-0.33$ s 
correspond to spatial scales $80-330$ km where non-MHD dissipation and damping processes are 
non-negligible \cite{bor03}. For what follows we will use the subscripts 'ss' for small scales
and 'ls' for large scales. 
Except for $\zeta_{ss}(q>0.5,B_X)$, $\zeta(q)$  exhibits 
linear dependence in the multiscale diagram (Figure 3). 
$\zeta_{ss}(q,B_X)$ is close to that linear dependence within $q\in(-1,0.5)$.

Figure 4 shows the scaling of moments during  interval B. $\zeta_{ls}(q)$ remains approximately linear
for both $B_X$ and $B_Z$. In comparison with  interval A, the small scale behavior is different, 
$\zeta_{ss}(q, B_X)\sim0$ for each $q$. In this case, the partition function (Eq.2)
does not represent the effects of intermittence adequately, because of the flat power spectrum
with a spectral index $\alpha=\zeta_{ss}(q=2,B_X)+1\sim1$ \cite{tu}. 
In contrast, $\zeta_{ss}(q<2,B_Z)$ follows
the straight line $\zeta(q)=q/3$, which describes the scaling in homogeneous Kolmogorov model of turbulence. 
However, $\zeta_{ss}(q>2,B_Z)$ becomes  undistinguishable from  the large scale scalings 
(within the confidence  intervals). This behavior indicates
a weak multifractality in small scale vertical fluctuations of the magnetic field. 
Here, two additional points
have to be clarified. First, in this paper we used magnetic field time series, 
therefore, when interpeting the
observed scalings in terms of turbulence models, we have to suppose the validity of the Taylor frozen field
hypothesis. In the plasma sheet the Taylor hypothesis is expected to be valid during fast BBFs \cite{hor}.
In our case plasma flow velocities achieve $\sim1500$ km/s during the chosen intervals and the validity of
the frozen field hypothesis seems to be substantiated. Second, instead 
of the velocity measurements, which are used in phenomenological models of hydrodynamic turbulence,
we have magnetic field measurements, therefore the interpretation of $\zeta_{ss}(q,B_Z)$ in terms
of Kolmogorov scaling might be difficult. Kolmogorov turbulence is completely described by its
velocity field. If a passive scalar field is subject to Kolmogorov turbulence, the resulting
scaling of the passive scalar field is also Kolmogorov \cite{cho}.  The small 
scale weak magnetic field in the plasma sheet can be moved as a passive scalar and 
its scaling then resembles that of the velocity field. During the chosen intervals A and B, 
the small scale  magnitude of the fluctuating magnetic field is  a few nT, while 
the small scale velocity fluctuations achieve 500 km/s. 

The 
significant difference between $\zeta_{ss}(q,B_X)$ and $\zeta_{ss}(q,B_Z)$ indicates that the 
small scale fluctuations
appear to be anisotropic. To check this we plotted the time evolution of the relative power 
$c_{fr}=c_f(B_Z)/c_f(B_X)$ at scales $0.08$ s and $5.4$ s in Figure 5. 
$c_{fr}$ was computed in the same way as $c_{f}$ in Figure 1, using sliding overlapping windows.  
While the large scale ($5.4$ s) relative power of the $B_Z$ and $B_X$ fluctuations is close to one
during the intervals A and B, the small scale ($0.08$ s) relative power shows significant enhancements.
It means that magnetic fluctuations are excited preferentially in vertical direction during the intervals
A and B. Outside of A and B the small scale power of the $B_X$ and $B_Z$ fluctuations is comparable. Large scale
magnetic fluctuations exhibit more power in $B_X$ than $B_Z$ also outside the intervals A and B. These
features show that the observed anisotropy is scale dependent and the occurrence of
BBFs can modify the preferable direction of fluctuations.

In summary, the main difference between the intervals A and B is that small-scale magnetic fluctuations 
are less homogeneous during the interval B. Turbulence characteristics were expected to
be influenced by the PSBL in A. Here the magnetic fluctuations can be more homogeneous  due to an 
effective mixing of the plasma. The mixing length ($ML$) in the turbulent plasma sheet can be computed
as the product of the integral time scale and the average root-mean-square velocity of turbulent flows, giving
$ML\sim 10000$ km \cite{bor97}. $ML$ is of the order of the average distance that turbulent eddies can travel
before colliding with each other. In Prandtl's mixing length theory \cite{lan} an effective 
viscosity is introduced which is proportional to $ML^2$ multiplied by the absolute value of the local
velocity gradient. The velocity gradients are of the same order during A and B (not shown). 
However, $ML$ is position 
dependent. For  turbulence close to a boundary, $ML$ rapidly decreases, which makes
the mixing of plasma more effective. Since the plasma sheet is thinner in A and thicker in B, $ML$
should be shorter in A than in B. Nevertheless, the large scale scaling exhibits the same homogeneity  
in both A and B. The spectral index can be estimated  as \cite{tu} 
$\alpha_{ls}=\zeta_{ls}(q=2)+1\sim2.7\pm0.7$.
Similar values were obtained in Ref. 5. Interestingly, in wall-bounded turbulent shear flows 
$\alpha\sim 1.6-2.2$ is found \cite{johan}. We think that the longer $ML$ led to the observed small 
scale anisotropy and weak multifractality during the interval B.

\section{\label{sec:level1}CONCLUSIONS}
Much of our recent knowledge about solar wind turbulence comes from  both spectral and 
non-Gaussian (higher order statistical) properties of multiscale fluctuations. It is well understandable, 
since the available range
of MHD scales of fluctations in the solar wind 
embraces more than six decades of wave number space. In contrast, the range of available MHD scales in 
the plasma sheet spans over less than two decades \cite{bor03}. In solar wind studies, depending on the 
length of the  time series,  moments ($q$-th order structure functions) up to  $q=20$ were computed
\cite{mars}. Similar studies were not accomplished for plasma sheet turbulence. 

In this paper we demostrated that a proper study of the turbulence in the plasma sheet 
requires a thorough knowledge of the underlying
non-steady physical conditions which can strongly influence the estimation of the turbulence
characteristics. We studied BBF-associated magnetic fluctuations under  conditions that
allowed to consider the  changing plasma sheet thickness and   finite size effects. 
Using unsupervised methods, finite size effects can lead to spurious estimations of the 
scaling characteristics in turbulence. A possible solution of this problem comes from the study 
of the scaling features of the empirical characteristic function at the origin. 
We have shown that, at least for the analysed events, statistical moments  
can at best be computed  for $q\in(-1,3)$. So, the range of the
available statistical moments is significantly different from that in  the solar wind. 

In spite of the restricted range,
we have found that 
when the plasma sheet is thinner (Interval A),
boundary effects lead to the shortening of the mixing length. The plasma is mixed more efficiently,
and the small scale fluctuations become more homogeneous. The large scale magnetic fluctuations
are not sensitive to the changes of the plasma sheet thickness. Both BBF-associated intervals exhibit the
same large scale scaling characteristics in the multiscale diagram similar to wall-bounded
turbulent shear flows.

We have shown that
non-homogeneous magnetic turbulence and a weak multifractality (nonlinear dependence
of $\zeta(q)$ on $q$ in multiscale diagram) develops in the vertical direction at small scales (0.08-0.33 s), 
when 
the fluctuations
occur in a thick plasma sheet (Interval B), 
far from the PSBL. Though the multifractal signatures are indicative of an
inhomogeneous energy transfer through a turbulent cascade, a model of intermittent  turbulence 
(e.g. the P-model) cannot be fitted to the nonlinear $\zeta(q)$, because in such a  model $\zeta(q=3)=1$ 
is expected \cite{pagela}. In our case, however,  $\zeta(q=3)>>1$. 

BBF associated magnetic fluctuations exhibit
multi-scale anisotropy features which are different from non-BBF periods. 
The  small scale scaling characteristics of $B_X$ and $B_Z$ fluctuations have found to be anisotropic in 
the multiscale diagrams.
The occurence of scale
dependent anisotropy is evident from the comparison  of the relative power of $B_Z$ and $B_X$
magnetic field fluctuations over two different scales ($0.08$ and $5.4$ s). 
Scale dependent anisotropy can robustly appear in MHD fluids in the 
presence of a local mean magnetic field \cite{biskamp}, but other mechanisms, e.g.  velocity shears 
can also produce  strong anisotropies \cite{ruder}.

A wider statistical study is needed, however, to explore 
fully the influence of the PSBL on magnetic turbulence and the appearance of anisotropy in the plasma sheet.

\begin{acknowledgments}
We thank H.-U. Eichelberger for help with FGM data.
\end{acknowledgments}

\pagebreak

\bibliography{voros}

\pagebreak
{\bf Figure captions}
\\
FIG. 1: Magnetic field measurements on Cluster 1 and 3; a. $B_X$ components; b. $B_Z$ components;
c. Small scale power of $B_Z$ fluctuations. \\
\\
FIG. 2: Scaling of the characteristic function (continuous line); the observed scaling regimes (dashed-dotted
lines); the exponent $\rho^+$ corresponds to the number of positive moments, $\rho^-$  
corresponds to the number of negative moments. \\
\\
FIG. 3: Multiscale diagram - scaling of the moments is depicted for $B_X$ and $B_Z$ components
at two different scale ranges within the interval A.  \\
\\
FIG. 4: Multiscale diagram - scaling of the moments is depicted for $B_X$ and $B_Z$ components
at two different scale ranges within the interval B. \\
\\
FIG. 5: Scale dependent anisotropy visible in time evolution of the relative power
$c_{fr}=c_f(B_Z)/c_f(B_X)$.

\pagebreak
\begin{figure}
\noindent
\includegraphics[width=20pc]{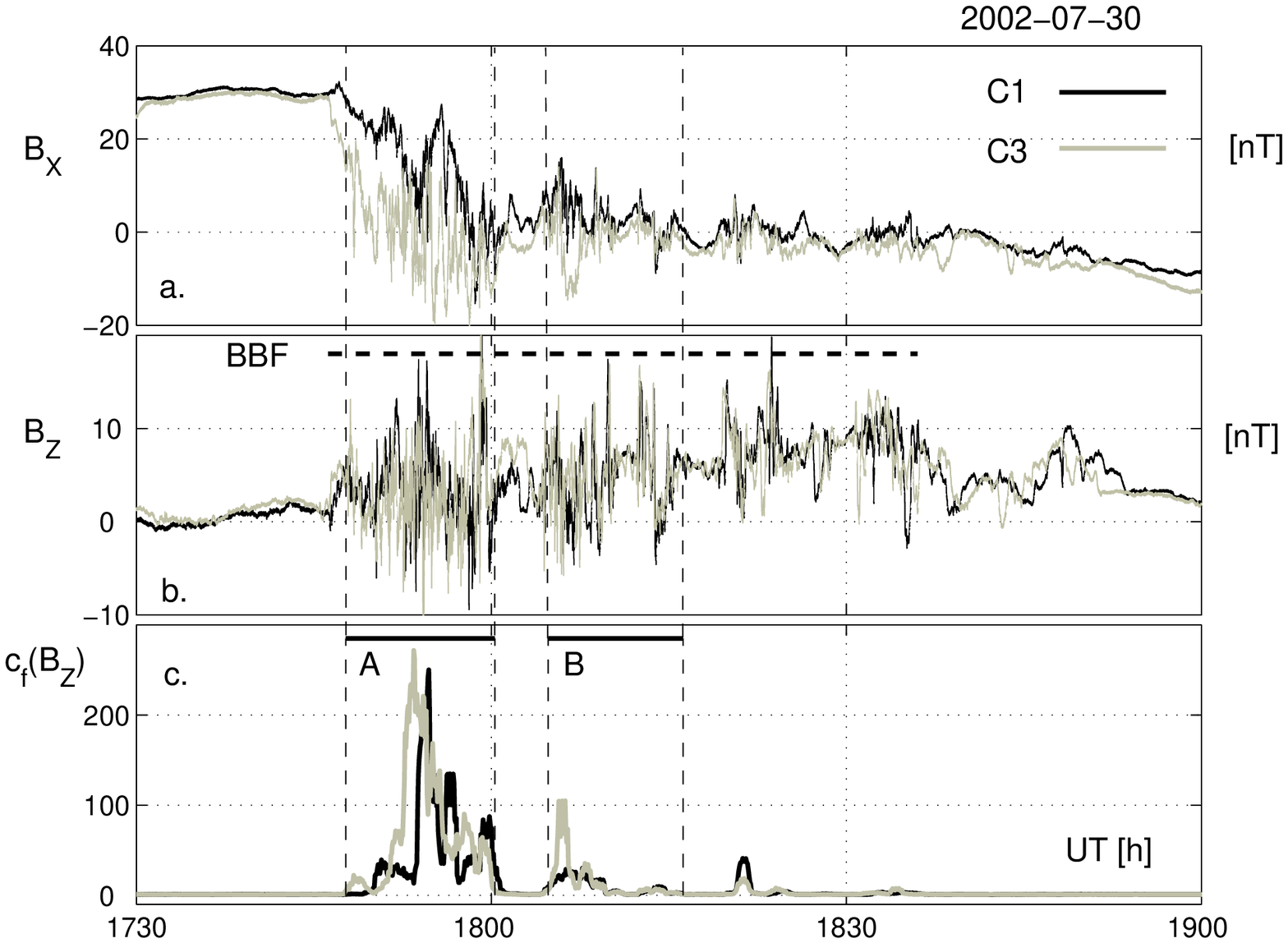}
\caption{}
\end{figure}

\begin{figure}
\noindent
\includegraphics[width=20pc]{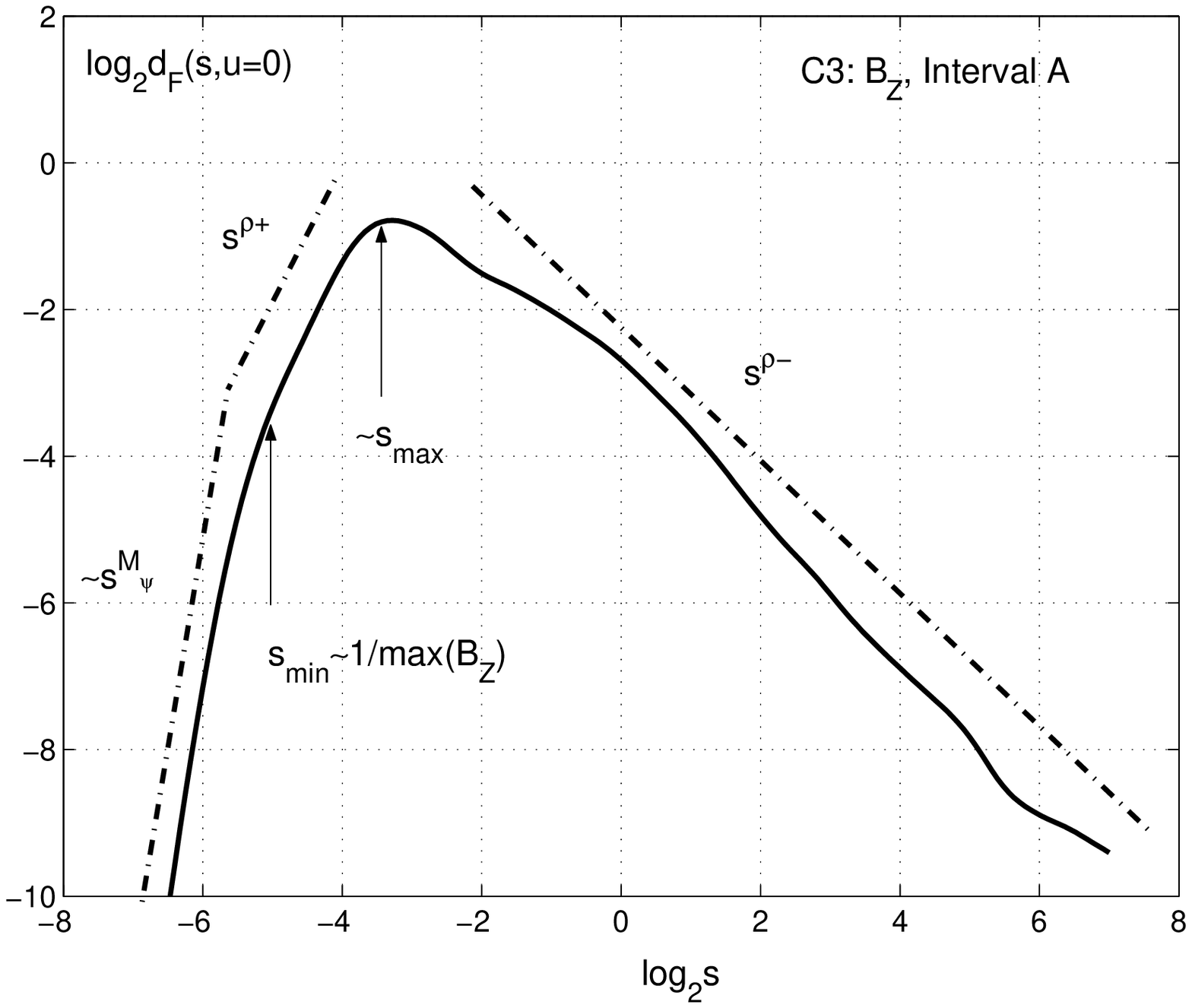}
\caption{}
\end{figure}

\begin{figure}
\noindent
\includegraphics[width=20pc]{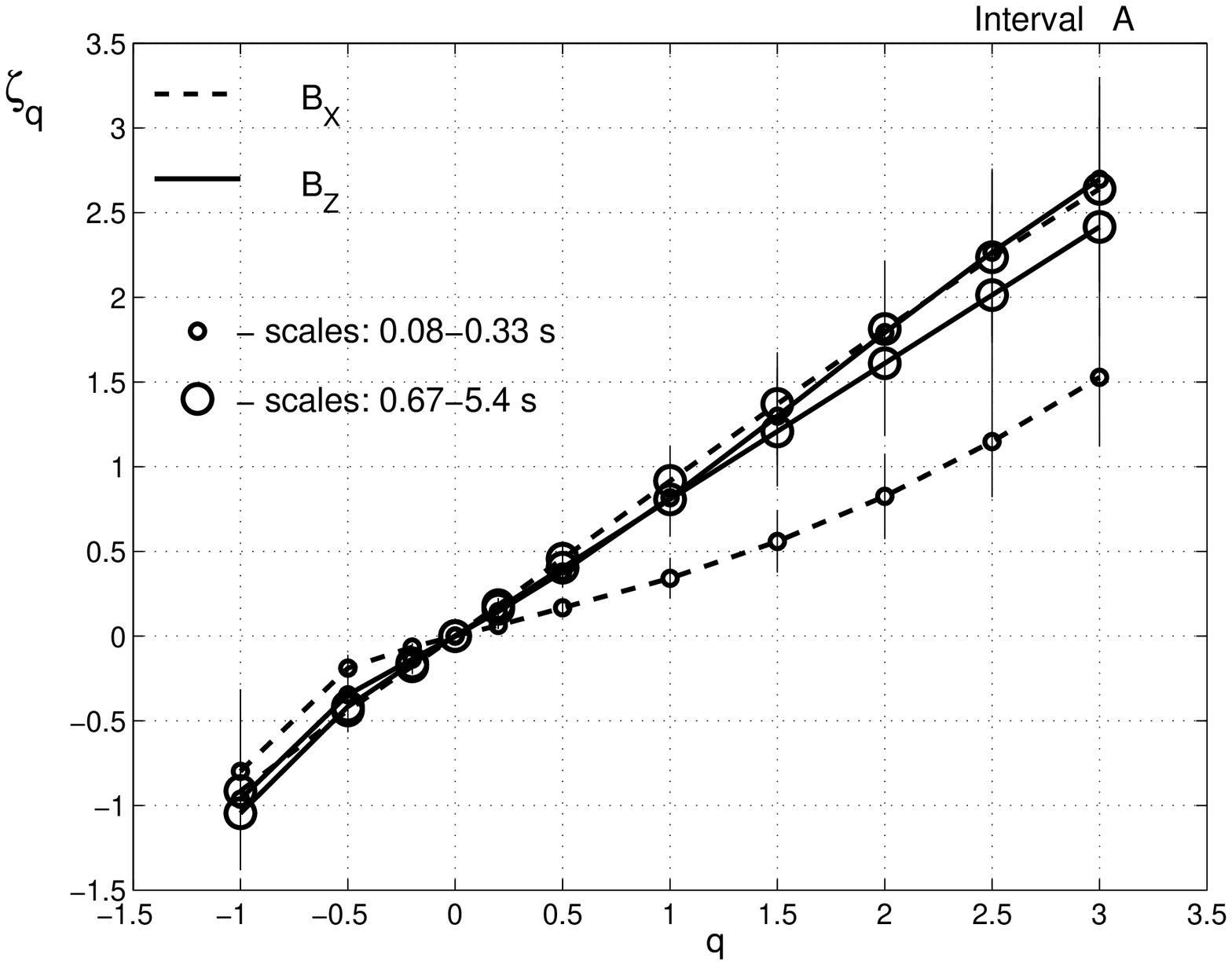}
\caption{}
\end{figure}

\begin{figure}
\noindent
\includegraphics[width=20pc]{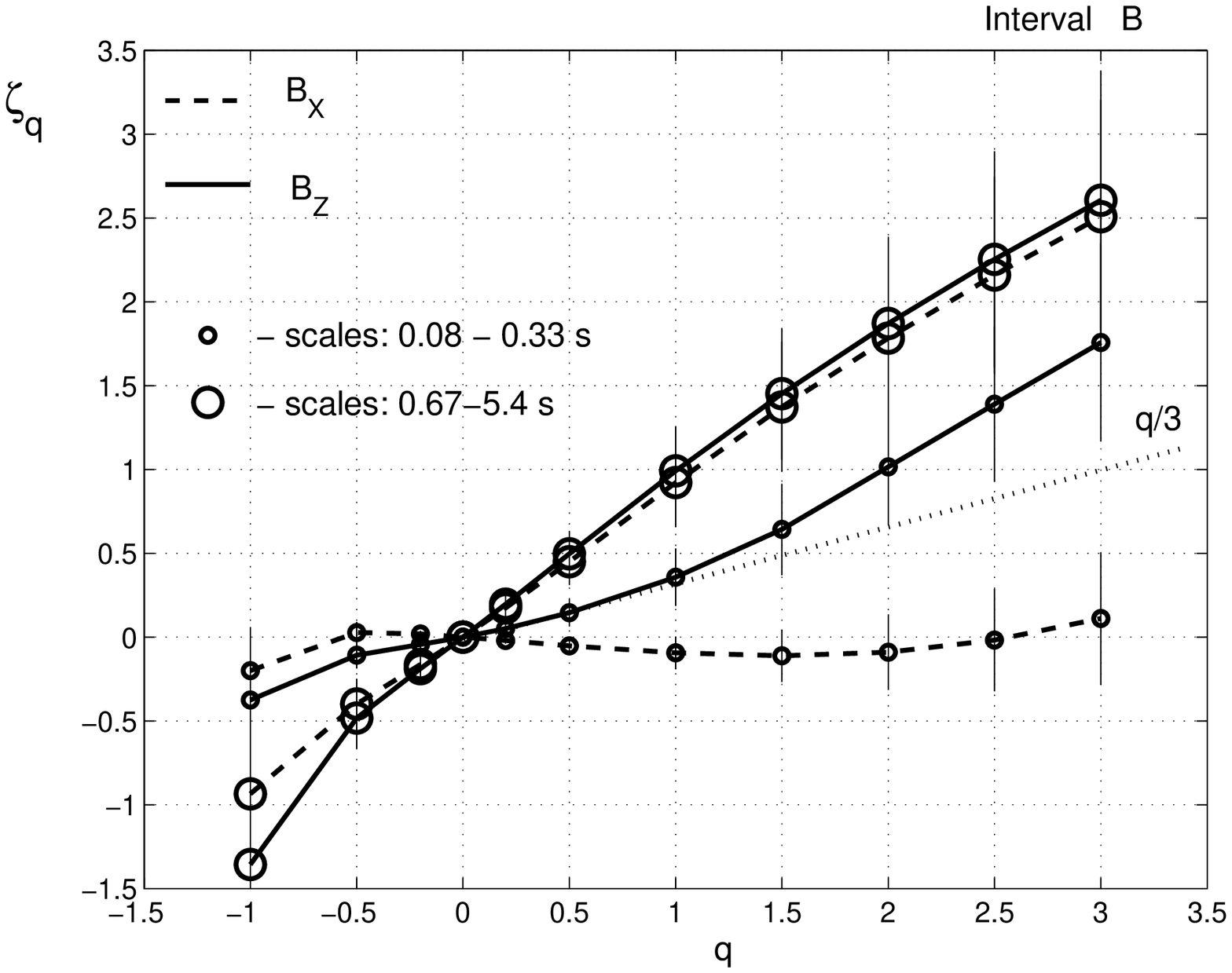}
\caption{}
\end{figure}

\begin{figure}
\noindent
\includegraphics[width=20pc]{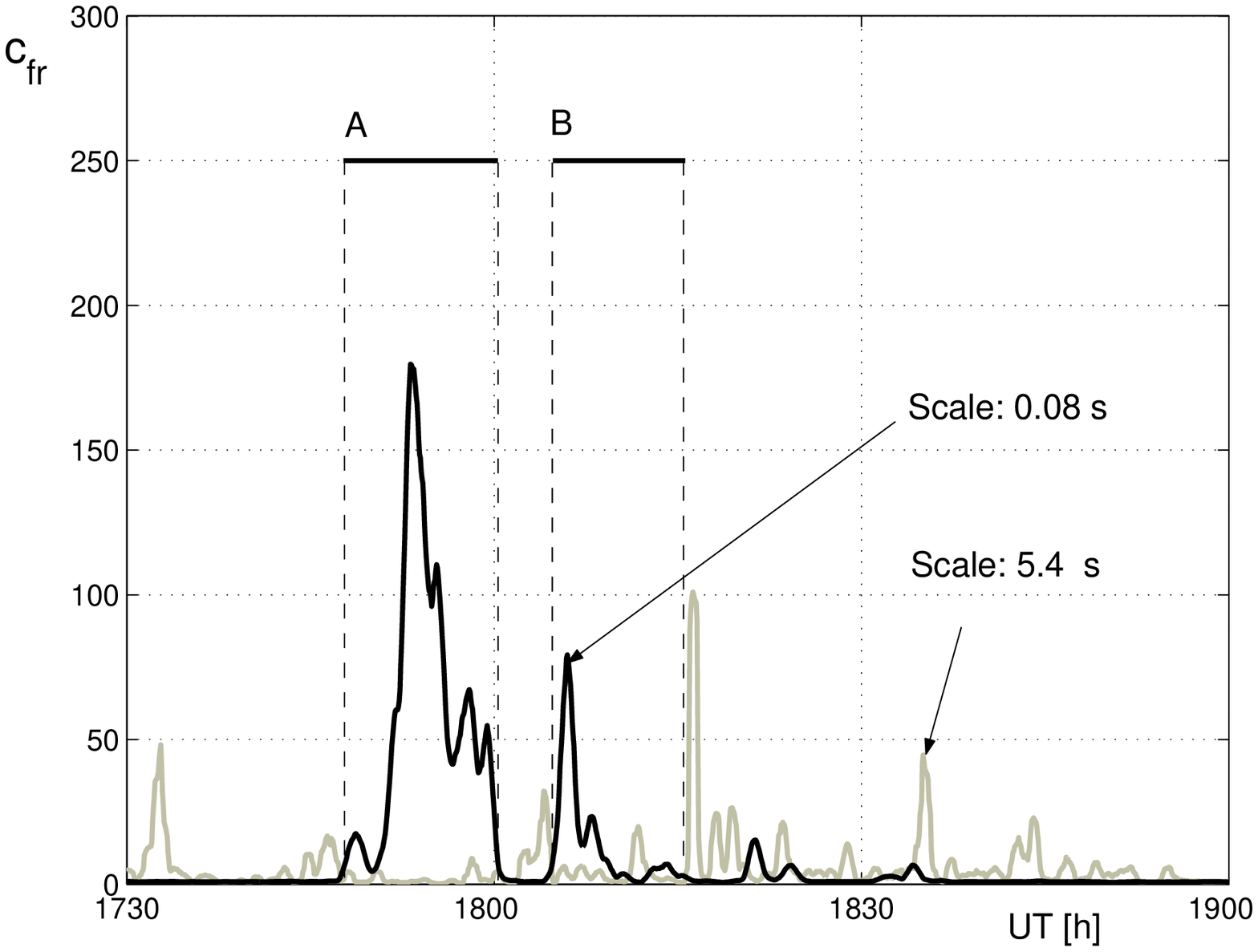}
\caption{}
\end{figure}

\end{document}